\documentclass[aps,prl,amsmath,twocolumn,amssymb,titlepage,showpacs,superscriptaddress]{revtex4-1}

\usepackage{graphicx}
\usepackage{nicefrac}
\usepackage{amsfonts}
\usepackage{amssymb}
\usepackage{amsmath} 
\usepackage{subfigure}
\usepackage{multirow} 
\usepackage{bm}
\usepackage{array}
\usepackage{verbatim}
\usepackage{url}
\usepackage{hyperref}

\newcommand{\ue}{\mathrm{e}}

\newcommand{\sgn}{\mathrm{sgn}}

\newcommand{\sfrac}[2]{\textstyle\frac{#1}{#2}\displaystyle}

\newcommand{\half}{\sfrac{1}{2}}

\begin{document}

\title{Kramers-Wannier Duality of Statistical Mechanics Applied to the Boolean Satisfiability Problem of Computer Science}
\author{Joe Mitchell}
\affiliation{Joint Quantum Institute and Condensed Matter Theory
  Center, Department of Physics, University of Maryland, College Park,
  Maryland 20742-4111, USA} 
\author{Benjamin Hsu}
\affiliation{Department of Physics, Princeton University, Princeton, NJ 08544}
\author{Victor Galitski}
\affiliation{Joint Quantum Institute and Condensed Matter Theory
  Center, Department of Physics, University of Maryland, College Park,
  Maryland 20742-4111, USA} 

\date{\today}

\begin{abstract}
We present a novel application of the Kramers-Wannier duality on one of the most
 important problems of computer science, the Boolean satisfiability problem (SAT).  More specifically,
 we focus on sharp-SAT or equivalently \#SAT -- the problem of counting the number of solutions to
 a Boolean satisfaction formula. \#SAT can be cast into a statistical-mechanical language, where it reduces
 to calculating the partition function of an Ising spin Hamiltonian with multi-spin interactions. We show that
 Kramers-Wannier duality can be generalized to apply to such multi-connected spin networks. We present an
 exact dual partner to \#SAT and explicitly verify their equivalence with a few simple examples. It is shown
 that the NP-completeness of the original problem maps on the complexity of the dual problem of enumerating
 the number of non-negative solutions to a Diophantine system of equations. We discuss the implications of
 this duality and the prospects of similar dualities applied to computer science problems.
\end{abstract}


\maketitle

Computationally intractable problems are ubiquitous and occur in many areas of the natural and computational
 sciences, with problems as diverse as spin glasses, optimization problems, and
 cryptography~\cite{AchlioptasHBSAT,Altarelli2008,Mezard2009,SebastianiHBSAT}.
 Generally speaking, they are problems whose solutions require a time that grows exponentially in the input
 size $N$ of the problem. Understanding what sorts of
 problems have solutions that require exponential or polynomial time is an important
 issue that has many practical implications ~\cite{Arora2009}. 

However, proving that a problem is indeed computationally intractable is by itself a difficult task. One tool
 utilized by theoretical computer scientists is mapping between problems with a known computational complexity.
 By mapping one problem into another, one can show their relative difficulty. In this work, we use for the first
 time the well-established statistical mechanics technique of Kramers-Wannier duality to analyze the famous Boolean
 satisfiability problem. We discover an a new and exact dual formulation of the counting problem \#SAT, which
 maps it on an under-constrained system of Diophantine equations.

Kramers and Wannier's work in 1941~\cite{Kramers1941} exactly related the partition function of the $2D$ square
 lattice Ising model to itself, deriving the exact transition temperature of the model.   The method consists of
 introducing new link variables, summing out the original variables, and defining the link variables in terms of
 plaquette variables in order to satisfy constraints  introduced by the summations. In this way, the plaquettes
 of the original lattice become the vertices of the dual lattice. The graphical dual of the square lattice is
 itself, so the square lattice Ising model is self-dual. With this relation, the knowledge that a phase transition
 must happen in the dual model when it happens in the regular model, and the assumption that there is only one
 phase transition in the square lattice Ising model, the transition temperature is easily derived.  This value
 was later confirmed by Onsager's exact solution for the model \cite{Onsager1944}.

Since then, dualities of this sort have been generalized to a broad range of models \cite{Savit1980,Kogut1979}.
 Furthermore, the recently invented bond algebraic dualities~\cite{Nussinov2009,Cobanera2010,Cobanera2011,Ortiz2012}
 have not only grouped many of these classical and quantum dualities under a general framework, but also present
 methods to discover new dualities. The duality we use in this work is essentially the classic Kramers-Wannier
 duality for an Ising model in an external field. This additional field changes the constraints introduced so that
 the variables and interactions switch under duality, rather than the variables switching with the plaquettes.
 This is a lattice-independent transformation and allows us to apply duality not only for models with multi-spin
 interactions, but also for networks of spins, with no lattice structure required.

We will focus our attention on the Boolean satisfiability problem in computer science, known as SAT.
 Let $\{x_i\}$ be a set of $N$ Boolean variables.  Define a ``clause'' as the logical OR ($\vee$) of some
 combination of these variables, possibly with negation, e.g. $(x_1 \vee \bar{x}_2 \vee x_5)$.
 A SAT problem asks whether some set of $M$ clauses of the $N$ Boolean variables may be simultaneously
 satisfied by some assignment of the variables. 
 SAT problems that have the same number $k$ of variables in each clause are known as $k$-SAT problems.

This problem is extremely important in computational complexity theory and
 computer science in general.  It is in the non-deterministic polynomial time computational complexity
 class (NP), meaning it is verifiable, but not necessarily solvable, in polynomial time.  It is in fact
 an NP-complete problem, meaning that in addition to being in NP itself, it is also as difficult as all
 other problems in NP \cite{Cook1971,Levin1973}.  A polynomial time
 solution for this problem would imply a polynomial time solution for all NP problems, answering the
 famous question P$\overset{?}{=}$NP.
 The SAT problem is actually the first proven example of an NP-complete problem,
 and many other NP-complete problems were proven so by reducing them to the SAT problem.

It is straightforward to examine
 the SAT problem from a statistical mechanics standpoint.  Consider that the Boolean variables are
 equivalent to Ising variables and that the clauses can be expressed as Hamiltonian contributions
 that are zero if satisfied and positive otherwise, represented as a combination of multispin
 interactions.  Many methods used to study sets of random SAT problems are also used to study
 disordered Ising models, including the replica method, belief propogation, and survey propogation~\cite{Mezard1987,Mezard2002,Mezard2009}.

Random $k$-SAT problems with a large number of variables and clauses have a variety of interesting behavior
 and have been a topic of study in statistical mechanics for some time. It was found for instance that in
 random $3$-SAT problems there is a phase transition at the critical ratio $\alpha_{c} = M/N \approx 4.25$.
 The majority of 3-SAT instances are satisfiable below $\alpha_c$, and unsatisfiable above $\alpha_c$.
 Moreover, the time to find a solution is polynomial in $N$ below the transition and exponential above~\cite{Mezard2002,Mezard2009}.
 At the threshold, there is an exponentially sharp peak in the median running time for the best
 known 3-SAT solvers to decide if there is a solution or not. From a statistical physics point of view, this
 exponential slow down corresponds to a spin glass transition in an equivalent disordered Ising model~\cite{Monasson1997,Montanari2008}.
 One of the primary motivations for studying duality in $k$-SAT is because of the close connection
 between phase transitions and duality. We hope that the duality derived here and similar dualities
 for other computer science problems may help study the phases and phase transitions in these problems,
 particularly in establishing exact relations between the phases of seemingly unrelated problems.

To be specific, in this article we are concerned with calculating the number of assignments of the Boolean
 variables that make the set of clauses true.  This problem
 is known as sharp-SAT, or \#SAT.
 While the ``counting problem'', \#SAT, is closely related to the ``decision problem'', SAT,
 they are distinct.  SAT is in the
 class of NP-complete problems; it is in NP and is as difficult as any other problem in NP.
 \#SAT is in the \#P-complete problems, where \#P roughly is the class of counting problems associated
 with decision problems that are in NP.  Because Kramers-Wannier duality is a relation between
 the partition functions of two problems and not, for example, between low energy states, we will be
 able to more easily investigate counting problems than decision problems.

A SAT instance can be mapped into a spin glass model. Define the Ising spin variable as
 $\sigma_{s} = (-1)^{x_{s}}$ where $x_{s} = 0,1$ is the Boolean variable in a given clause, and define
 $c_s^\mu$ to be $-1$ if $x_s$ is involved in clause $\mu$ with a negation, $1$ if it is involved
 without a negation, and $0$ if it is not involved in the clause.
 One can define an energy cost for a violated clause as proportional to a number $X_\mu>0$ with the Hamiltonian
\begin{equation}\label{eqn:model}
  \mathcal{H} = \sum_\mu X_\mu \left[ \prod_s (1 - c_s^\mu \sigma_s) \right].
\end{equation}
It is easy to see that $\mathcal{H}\geq0$.  The decision problem is equivalent to determining if
 this Hamiltonian has any ground states with zero energy.  The counting problem is equivalent to
 calculating how many states have zero energy.  This can be compactly described by calculating
 the associated partition function, $\mathcal{Z}=\sum_{\{\sigma_s\}} \ue^{-\mathcal{H}}$, in the limit
 as each $X_\mu\to\infty$.

This Hamiltonian can be rearranged to suggest its nature as an Ising spin glass:
\begin{equation}
\mathcal{H} = \mathcal{H}_0 - \sum_s H_s \sigma_s - \sum_l T_l \prod_{s\in\partial l} \sigma_s.
\end{equation}
\begin{eqnarray}
\mathcal{H}_0 = \sum_\mu  X_\mu,
\qquad
H_s = \sum_\mu X_\mu c_s^\mu,
\\\notag
T_l = - \sum_\mu X_\mu \prod_{s\in\partial l} (-c_s^\mu).
\end{eqnarray}
Note that each of $\mathcal{H}_0$, $H_s$, and $T_l$ are linearly proportional to the large constants $X_\mu$.
 This is the model of a network of Ising spins with multispin interactions
 and a site-dependent magnetic field.  E.g.\ in the case of $3$-SAT, there are two- and three-body interactions.
 In this article, $s$ will signify the ``sites'' or locations of the variables $\sigma_s$,
 and $l$ will signify the ``links'' or locations of the multispin interactions $\prod_{s\in\partial l} \sigma_s$.
 $\partial l$ refers to the sites of spins involved in the interaction at $l$, and $\partial s$ refers to
 the links for interactions in which the spin $\sigma_s$ is involved.
 
\begin{figure}[htbp]
  \centering
  \includegraphics[width=\columnwidth]{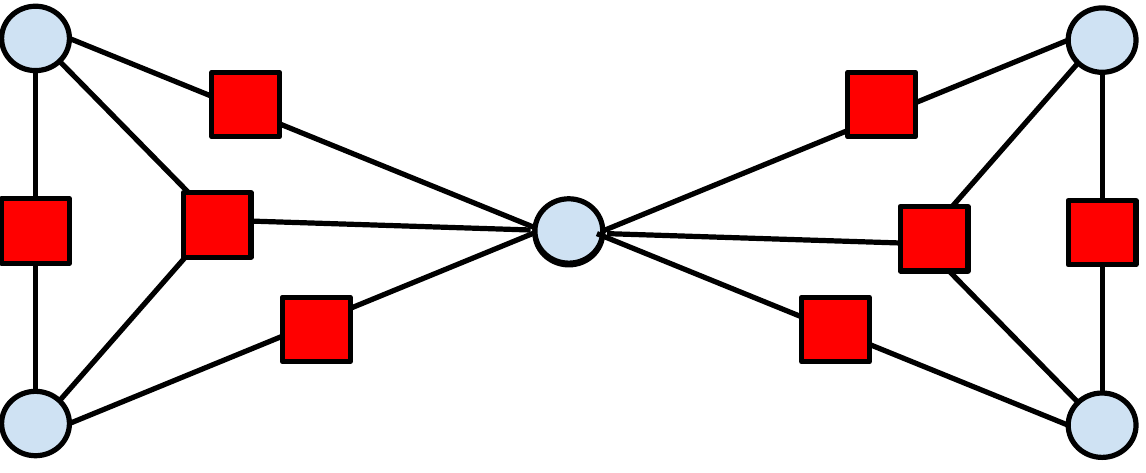}
  \caption{A SAT instance, $\mathcal{F}=(x_1 \vee x_2 \vee x_3) \wedge (x_3 \vee x_4 \vee x_5)$.
    Each circle represents a site, the location of a spin associated with a Boolean variable.
    Each square represents a link, the location of an interaction (constraint) between spins.
    In the dual picture, the spins are exchanged with interactions.\label{fig:example} }
\end{figure}

For any model of this type, there is a relatively simple Kramers-Wannier like duality~\cite{Savit1980,Supplement}.
\begin{equation}
\mathcal{Z} = \ue^{-\mathcal{H}_0} \sum_{\{\sigma_s=\pm1\}} \ue^{\sum_s H_s \sigma_s + \sum_l T_l \prod_{s\in\partial l} \sigma_s}
 = \mathcal{Z}_0 \tilde{\mathcal{Z}},
\end{equation}
\begin{equation}
\mathcal{Z}_0 = \ue^{-\mathcal{H}_0} \left( \prod_s 2 \sinh 2H_s \right)^{\frac12} \left( \prod_l \half \sinh 2T_l \right)^{\frac12}
\end{equation}
\begin{equation}
\tilde{\mathcal{Z}} = \sum_{\{\nu_l=\pm1\}} \ue^{\sum_l \tilde{T}_l \nu_l + \sum_s \tilde{H}_s \prod_{l\in\partial s} \nu_l},
\end{equation}
\begin{equation}
\tilde{H}_s = -\half \ln \tanh H_s,
\qquad
\tilde{T}_l = -\half \ln \tanh T_l.
\end{equation}
In short, dual spins are introduced at the links of the original network and the original spins are summed
 out, leaving interactions for the dual spins in their places.  I.e. the sites and links switch under duality.
 In addition, the dual couplings are monotonically decreasing functions of the original, and if the original
 couplings are negative then the dual couplings give a negative Boltzmann weight.  The effect of a complex
 coupling in the Ising model \cite{Matveev2008} when dealing with duality could be of some interest, but the
 SAT problem has real couplings and we do not consider it here.

Now we will simplify using the specifics of the SAT model.
 We must take the limit $X_\mu\to\infty$ prudently, Taylor expanding the partition function in terms of
 $\ue^{X_\mu}$.  With careful work, the partition function becomes:
\begin{equation} \label{eqn:full-dual}
\mathcal{Z} = \sum_{k_s,m_l \in \mathbb{Z}_+} \ue^{-\mathcal{S}_{\{k,m\}}}
 (-1)^{\sum_s k_s + \sum_l m_l} \times z_{\{k,m\}}
\end{equation}
\begin{equation}\label{eqn:z-alpha}
z_{\{k,m\}} = \sum_{\{\alpha_s\}} \left(\prod_s D_{k_s}^{\sgn H_s \alpha_s}\right)
 \left(\prod_l D_{m_l}^{\sgn T_l \prod_{s\in\partial l} \alpha_s}\right),
\end{equation}
\begin{equation}
D_k^{\alpha} = \half (1 + \alpha) - \half (1 - \alpha) (1-\delta_{k,0}),
\end{equation}
\begin{align}\label{eqn:action}
\mathcal{S}_{\{k,m\}} = \sum_\mu X_\mu \bigg[ 1 - &\sum_s (1-2k_s) \sgn H_s c_s^\mu
\\\notag
&- \sum_l (1-2m_l) \sgn T_l \prod_{s\in\partial l} c_s^\mu \bigg].
\end{align}
This does not seem a particularly enlightening form of the partition function, or of the \#SAT problem.
 We have introduced extra summations over the natural numbers for each site and link.  However, there are
 two points that should be considered.

First, consider the $X_\mu$ independent part of the partition function, Eq.~\eqref{eqn:z-alpha}.
 Given a set of $\{k_s,m_l\}$, this calculation is much simpler than it appears.
 Because of the particular simplicity of $D_k^\alpha$, an algorithm may be designed to
 calculate the $\alpha_s$ summation in time polynomial in $N$.  Effectively, this is an unimportant 
 calculation when compared to the full \#SAT.

The second point to consider is that the limit $X_\mu\to\infty$ will greatly restrict the summation over
 $k_s$ and $m_l$.  Since we control each $X_\mu$, we can set them so that $X_\mu$ is much greater than
 $X_{\mu+1}$.  Given this, $\sgn H_s = c_s^\mu$ in the \emph{first} clause where $c_s^\mu$ is nonzero, and
 $\sgn T_l = -\prod_{s\in\partial l} (-c_s^\mu)$ in the first clause where the product is nonzero.  Now
 consider $S_{\{k,m\}}$, a sum of terms proportional to the $X_\mu$.  If this sum is positive, then the
 contribution to the partition function vanishes.  If the
 sum is negative, then the sums must arrange to cancel this term or else the partition function
 would diverge.  The only relevant sets of $\{k,m\}$ in the summation are those that give $\mathcal{S}_{\{k,m\}}=0$.

Each $X_\mu$ is controlled separately, so we can set each term of $\mathcal{S}$ to zero separately:
\begin{equation} \label{eqn:Diophantine}
A^\mu = \sum_s B_s^\mu k_s + \sum_l C_l^\mu m_l,
\end{equation}
\begin{eqnarray}
&B_s^\mu = \sgn H_s c_s^\mu,
\qquad
C_l^\mu = - \sgn T_l \prod_{s\in\partial l} (-c_s^\mu),
\\\notag
&A^\mu = -\half\left( 1 - \sum_s B_s^\mu - \sum_l C_l^\mu \right).
\end{eqnarray}
This is a system of linear Diophantine equations.  Finding the nonnegative $k_s$ and $m_l$ that satisfy
 these equations gives exactly the $\{k,m\}$ that should be summed over in the partition function,
 Eq.~\eqref{eqn:full-dual}.  The complexity of calculating the number
 of solutions to a Boolean satisfaction instance has transformed into finding the nonnegative solutions
 of this (likely) underconstrained system of equations and summing an integer function over these solutions.

Eqs.~\eqref{eqn:full-dual}-\eqref{eqn:Diophantine} are the main results of this article.  We have shown
 that a problem in \#SAT is equivalent to listing the solutions to this system of integer equations.
 It would be useful to simplify these equations to also find an equivalent for the decision problem,
 i.e. whether $\mathcal{Z}=0$ or not.  However, this is no trivial task.  Any given
 solution to Eq.~\eqref{eqn:Diophantine} could give a positive or negative contribution to the partition
 function.  Knowing a portion of the solutions may give little information, as the rest may cancel the
 contribution to the partition function from the portion known, and so it will be difficult to say whether an
 instance is satisfiable or not.  This may change if solutions can be categorized or if the important
 solutions can be singled out, but naively only \#SAT is considered.

Now we will show simple examples of this duality.  First, we take a look at a single clause of three
 variables:
\begin{equation}
\mathcal{F} = (x_1 \vee x_2 \vee x_3).
\end{equation}
This equation is satisfiable, and has $\mathcal{Z}=7$ solutions that satisfy $\mathcal{F}=T$, all possible
 configurations except $x_1=x_2=x_3=F$.  Now we will use \eqref{eqn:full-dual} and \eqref{eqn:Diophantine}
 to calculate this number.

First, the sites are $s=1,2,3$ and the links are $l=12,13,23,123$.  Seeing that $c_s^\mu=1$,
 we find $\sgn H_s=\sgn T_{123}=1$ and for all $l \neq 123$, $\sgn T_l=-1$.  This means the
 Diophantine system of equations is just
\begin{equation}
k_1 + k_2 + k_3 + m_{12} + m_{13} + m_{23} + m_{123} = 3.
\end{equation}
The partition function is
\begin{equation}
\mathcal{Z} = \sum_{\{k_s,m_l \text{ s.t. } \mathcal{S}_{\{k,m\}}=0 \}} (-1)^{\sum_s k_s + \sum_l m_l} z_{\{k,m\}}
\end{equation}
Using \eqref{eqn:z-alpha}, $z_{\{k,m\}}$ may be calculated with
 relative simplicity.  There are eight configurations of $\{\delta_{k_s,0},\delta_{m_l,0}\}$ that leave
 $z$ nonzero.  One of these requires all seven of the $k$ and $m$ to be nonzero, while the others
 require exactly three nonzero and four zero.  These each contribute $1$ to the partition function,
 leaving us with $\mathcal{Z}=7$.

Now look at a very simple unsatisfiable case:
\begin{align}
\mathcal{F} = &(x_1 \vee x_2 \vee x_3) \wedge (x_1 \vee x_2 \vee \overline{x}_3)
\\\notag
& \wedge (x_1 \vee \overline{x}_2 \vee x_3) \wedge (x_1 \vee \overline{x}_2 \vee \overline{x}_3) \wedge (\overline{x}_1 \vee x_2 \vee x_3)
\\\notag
&\wedge (\overline{x}_1 \vee x_2 \vee \overline{x}_3) \wedge (\overline{x}_1 \vee \overline{x}_2 \vee x_3) \wedge (\overline{x}_1 \vee \overline{x}_2 \vee \overline{x}_3)
\end{align}
First, note that the calculation of $z_{\{k,m\}}$ is exactly the same.  When calculating
 a new \#SAT case, $z_{\{k,m\}}$ need not be recalculated unless the sites or links are changed.
 A simple benefit from this is that it makes it easier to analyze all bit flips of a SAT instance.
 However, the $\{k,m\}$ that give $\mathcal{S}=0$ must be recalculated each time.  In this case,
 the Diophantine system is overconstrained and unsatisfiable.  This immediately gives $\mathcal{Z}=0$.

An overconstrained system is one way in which SAT instance may reveal its unsatisfiability.
 However, this is not the only way.  As noted previously, the contributions to the partition function,
 $(-1)^{\sum_s k_s + \sum_l m_l} z_{\{k,m\}}$, may be positive or negative, and it is possible
 for there to be many solutions to the Diophantine system that give contributions that cancel.
 For another example and more discussion on unsatisfiability,
 see section 2 of the supplement~\cite{Supplement}.



It is important to ask how difficult it is to find the solutions to Eq.~\eqref{eqn:Diophantine}.
 First, let us discuss the matrix size.  Although the equations apply to any
 instance, let us restrict to interesing instances,
 with a large number of variables and clauses, roughly of the same
 order:  $N \sim M >> 1$.  Let us also assume the clauses contain a small number of variables.
 Assuming the clauses are well distributed among the variables,
 the matrix is roughly of the size $N+kM \times M$, where $k$ is some small number related to the
 number of variables in the clauses.  The matrix size is quadratic in input size $N$.

There is much research addressing
 algorithms for reducing a Diophantine system to useable forms, in particular the Hermite
 and Smith normal forms~\cite{Mulders1999,Jager2009}.  The algorithms take more time than the analagous reduced
 row echelon algorithms for real matrices but the time is still polynomial in $N$.
 However, acquiring the nonnegative solutions to the system is not a polynomial task.
 Indeed, the number of nonnegative solutions need not be
 polynomial in $N$.  For example, a collection of $M$ clauses with $3$ variables where no
 clauses share variables will have $7^M$ solutions.  Simply listing the solutions is \#P-hard.


There is no doubt that there are instances where $\mathcal{Z}$ is easier to calculate
 without using duality.  However, it is quite possible that there are situations where it
 will be easier to use the dual equations, notably when the number of solutions
 to the Diophantine system is small or zero.  This situation is likely to coincide to some
 extent with when the number of solutions to SAT instances is small or zero,
 the region of greatest interest in Boolean satisfiability research.

Beyond this, the particular form of the matrix (the fact that it is a $(-1,0,1)$
 matrix and how the link columns are completely determined from the site columns) could
 lead to great simplification in algorithms for finding the nonnegative solutions.
 Also, by categorizing which solutions will give negative values, the equations could
 be simplified to describe the decision problem which is of more interest, SAT itself.

Combining duality with the cavity method, commonly used to study SAT problems as well
 as Ising problems, could be a promising research direction.  Preliminary steps
 may be found in the supplement~\cite{Supplement}.


Other computer science problems could also have interesting dual problems.  Many
 of these problems are easily stated as some limit of a statistical mechanics
 model, which likely has a simple duality relation to
 another model.  Once the appropriate limit is taken, this dual problem
 will reduce to a problem very unlike the original.  The $x-y$, $p$-clock,
 Coulomb gas, and vector Potts models are a small set of classical models with well
 studied, simple duality relations~\cite{Savit1980,Kogut1979}.


The recent work with bond algebraic dualities \cite{Nussinov2009,Cobanera2010,Cobanera2011,Ortiz2012}
 opens up the number
 of models to study even further.  With this new way to categorize and explore
 dualities, even quantum complexity classes could be accessible for study with
 duality~\cite{Laumann2010,Hsu2013}.

In this letter, we have presented an exact relation between problems in \#SAT
 and solving a linear Diophantine system of equations, given from a modified version
 of the established Kramers-Wannier duality.  This relation serves both as a novel
 avenue for study of \#SAT and a glimpse at the power statistical mechanics
 dualities can bring to computer science.  With the large number of duality relations
 and the ease of relating a computer science problem to the limit of a physical
 model, the study of duality could prove very fruitful.

This research was supported by NSF-CAREER award No. DMR-0847224 and Simons Foundation.

\bibliography{SATDual}

\begin{thebibliography}{26}%
\makeatletter
\providecommand \@ifxundefined [1]{%
 \@ifx{#1\undefined}
}%
\providecommand \@ifnum [1]{%
 \ifnum #1\expandafter \@firstoftwo
 \else \expandafter \@secondoftwo
 \fi
}%
\providecommand \@ifx [1]{%
 \ifx #1\expandafter \@firstoftwo
 \else \expandafter \@secondoftwo
 \fi
}%
\providecommand \natexlab [1]{#1}%
\providecommand \enquote  [1]{``#1''}%
\providecommand \bibnamefont  [1]{#1}%
\providecommand \bibfnamefont [1]{#1}%
\providecommand \citenamefont [1]{#1}%
\providecommand \href@noop [0]{\@secondoftwo}%
\providecommand \href [0]{\begingroup \@sanitize@url \@href}%
\providecommand \@href[1]{\@@startlink{#1}\@@href}%
\providecommand \@@href[1]{\endgroup#1\@@endlink}%
\providecommand \@sanitize@url [0]{\catcode `\\12\catcode `\$12\catcode
  `\&12\catcode `\#12\catcode `\^12\catcode `\_12\catcode `\%12\relax}%
\providecommand \@@startlink[1]{}%
\providecommand \@@endlink[0]{}%
\providecommand \url  [0]{\begingroup\@sanitize@url \@url }%
\providecommand \@url [1]{\endgroup\@href {#1}{\urlprefix }}%
\providecommand \urlprefix  [0]{URL }%
\providecommand \Eprint [0]{\href }%
\providecommand \doibase [0]{http://dx.doi.org/}%
\providecommand \selectlanguage [0]{\@gobble}%
\providecommand \bibinfo  [0]{\@secondoftwo}%
\providecommand \bibfield  [0]{\@secondoftwo}%
\providecommand \translation [1]{[#1]}%
\providecommand \BibitemOpen [0]{}%
\providecommand \bibitemStop [0]{}%
\providecommand \bibitemNoStop [0]{.\EOS\space}%
\providecommand \EOS [0]{\spacefactor3000\relax}%
\providecommand \BibitemShut  [1]{\csname bibitem#1\endcsname}%
\let\auto@bib@innerbib\@empty
\bibitem [{\citenamefont {Achlioptas}(2009)}]{AchlioptasHBSAT}%
  \BibitemOpen
  \bibfield  {author} {\bibinfo {author} {\bibfnamefont {D.}~\bibnamefont
  {Achlioptas}},\ }\enquote {\bibinfo {title} {Random satisfiability},}\
  Chap.~\bibinfo {chapter} {8}, pp.\ \bibinfo {pages} {245--270},\ vol.\
  \bibinfo {volume} {185}\ of\  \cite{Biere2009} (\bibinfo {year}
  {2009})\BibitemShut {NoStop}%
\bibitem [{\citenamefont {Altarelli}\ \emph {et~al.}(2008)\citenamefont
  {Altarelli}, \citenamefont {Monasson}, \citenamefont {Semerjian},\ and\
  \citenamefont {Zamponi}}]{Altarelli2008}%
  \BibitemOpen
  \bibfield  {author} {\bibinfo {author} {\bibfnamefont {F.}~\bibnamefont
  {Altarelli}}, \bibinfo {author} {\bibfnamefont {R.}~\bibnamefont {Monasson}},
  \bibinfo {author} {\bibfnamefont {G.}~\bibnamefont {Semerjian}}, \ and\
  \bibinfo {author} {\bibfnamefont {F.}~\bibnamefont {Zamponi}},\ }\href@noop
  {} {\bibfield  {journal} {\bibinfo  {journal} {CoRR}\ }\textbf {\bibinfo
  {volume} {abs/0802.1829}} (\bibinfo {year} {2008})}\BibitemShut {NoStop}%
\bibitem [{\citenamefont {M{\'e}zard}\ and\ \citenamefont
  {Montanari}(2009)}]{Mezard2009}%
  \BibitemOpen
  \bibfield  {author} {\bibinfo {author} {\bibfnamefont {M.}~\bibnamefont
  {M{\'e}zard}}\ and\ \bibinfo {author} {\bibfnamefont {A.}~\bibnamefont
  {Montanari}},\ }\href@noop {} {\emph {\bibinfo {title} {{Information,
  physics, and computation}}}}\ (\bibinfo  {publisher} {Oxford University Press
  Inc.},\ \bibinfo {address} {New York, NY, USA},\ \bibinfo {year}
  {2009})\BibitemShut {NoStop}%
\bibitem [{\citenamefont {Sebastiani}\ and\ \citenamefont
  {Tacchella}(2009)}]{SebastianiHBSAT}%
  \BibitemOpen
  \bibfield  {author} {\bibinfo {author} {\bibfnamefont {R.}~\bibnamefont
  {Sebastiani}}\ and\ \bibinfo {author} {\bibfnamefont {A.}~\bibnamefont
  {Tacchella}},\ }\enquote {\bibinfo {title} {Sat techniques for modal and
  description logics},}\ Chap.~\bibinfo {chapter} {25}, pp.\ \bibinfo {pages}
  {781--824},\ vol.\ \bibinfo {volume} {185}\ of\  \cite{Biere2009} (\bibinfo
  {year} {2009})\BibitemShut {NoStop}%
\bibitem [{\citenamefont {Arora}\ and\ \citenamefont
  {Barak}(2009)}]{Arora2009}%
  \BibitemOpen
  \bibfield  {author} {\bibinfo {author} {\bibfnamefont {S.}~\bibnamefont
  {Arora}}\ and\ \bibinfo {author} {\bibfnamefont {B.}~\bibnamefont {Barak}},\
  }\href@noop {} {\emph {\bibinfo {title} {Complexity Theory: A Modern
  Approach}}}\ (\bibinfo  {publisher} {Cambridge University Press},\ \bibinfo
  {year} {2009})\BibitemShut {NoStop}%
\bibitem [{\citenamefont {Kramers}\ and\ \citenamefont
  {Wannier}(1941)}]{Kramers1941}%
  \BibitemOpen
  \bibfield  {author} {\bibinfo {author} {\bibfnamefont {H.~A.}\ \bibnamefont
  {Kramers}}\ and\ \bibinfo {author} {\bibfnamefont {G.~H.}\ \bibnamefont
  {Wannier}},\ }\href {\doibase 10.1103/PhysRev.60.252} {\bibfield  {journal}
  {\bibinfo  {journal} {Phys. Rev.}\ }\textbf {\bibinfo {volume} {60}},\
  \bibinfo {pages} {252} (\bibinfo {year} {1941})}\BibitemShut {NoStop}%
\bibitem [{\citenamefont {Onsager}(1944)}]{Onsager1944}%
  \BibitemOpen
  \bibfield  {author} {\bibinfo {author} {\bibfnamefont {L.}~\bibnamefont
  {Onsager}},\ }\href {\doibase 10.1103/PhysRev.65.117} {\bibfield  {journal}
  {\bibinfo  {journal} {Phys. Rev.}\ }\textbf {\bibinfo {volume} {65}},\
  \bibinfo {pages} {117} (\bibinfo {year} {1944})}\BibitemShut {NoStop}%
\bibitem [{\citenamefont {Savit}(1980)}]{Savit1980}%
  \BibitemOpen
  \bibfield  {author} {\bibinfo {author} {\bibfnamefont {R.}~\bibnamefont
  {Savit}},\ }\href {\doibase 10.1103/RevModPhys.52.453} {\bibfield  {journal}
  {\bibinfo  {journal} {Rev. Mod. Phys.}\ }\textbf {\bibinfo {volume} {52}},\
  \bibinfo {pages} {453} (\bibinfo {year} {1980})}\BibitemShut {NoStop}%
\bibitem [{\citenamefont {Kogut}(1979)}]{Kogut1979}%
  \BibitemOpen
  \bibfield  {author} {\bibinfo {author} {\bibfnamefont {J.~B.}\ \bibnamefont
  {Kogut}},\ }\href {\doibase 10.1103/RevModPhys.51.659} {\bibfield  {journal}
  {\bibinfo  {journal} {Rev. Mod. Phys.}\ }\textbf {\bibinfo {volume} {51}},\
  \bibinfo {pages} {659} (\bibinfo {year} {1979})}\BibitemShut {NoStop}%
\bibitem [{\citenamefont {Nussinov}\ and\ \citenamefont
  {Ortiz}(2009)}]{Nussinov2009}%
  \BibitemOpen
  \bibfield  {author} {\bibinfo {author} {\bibfnamefont {Z.}~\bibnamefont
  {Nussinov}}\ and\ \bibinfo {author} {\bibfnamefont {G.}~\bibnamefont
  {Ortiz}},\ }\href {\doibase 10.1103/PhysRevB.79.214440} {\bibfield  {journal}
  {\bibinfo  {journal} {Phys. Rev. B}\ }\textbf {\bibinfo {volume} {79}},\
  \bibinfo {pages} {214440} (\bibinfo {year} {2009})}\BibitemShut {NoStop}%
\bibitem [{\citenamefont {Cobanera}\ \emph {et~al.}(2010)\citenamefont
  {Cobanera}, \citenamefont {Ortiz},\ and\ \citenamefont
  {Nussinov}}]{Cobanera2010}%
  \BibitemOpen
  \bibfield  {author} {\bibinfo {author} {\bibfnamefont {E.}~\bibnamefont
  {Cobanera}}, \bibinfo {author} {\bibfnamefont {G.}~\bibnamefont {Ortiz}}, \
  and\ \bibinfo {author} {\bibfnamefont {Z.}~\bibnamefont {Nussinov}},\ }\href
  {\doibase 10.1103/PhysRevLett.104.020402} {\bibfield  {journal} {\bibinfo
  {journal} {Phys. Rev. Lett.}\ }\textbf {\bibinfo {volume} {104}},\ \bibinfo
  {pages} {020402} (\bibinfo {year} {2010})}\BibitemShut {NoStop}%
\bibitem [{\citenamefont {Cobanera}\ \emph {et~al.}(2011)\citenamefont
  {Cobanera}, \citenamefont {Ortiz},\ and\ \citenamefont
  {Nussinov}}]{Cobanera2011}%
  \BibitemOpen
  \bibfield  {author} {\bibinfo {author} {\bibfnamefont {E.}~\bibnamefont
  {Cobanera}}, \bibinfo {author} {\bibfnamefont {G.}~\bibnamefont {Ortiz}}, \
  and\ \bibinfo {author} {\bibfnamefont {Z.}~\bibnamefont {Nussinov}},\
  }\href@noop {} {\bibfield  {journal} {\bibinfo  {journal} {Advances in
  Physics}\ }\textbf {\bibinfo {volume} {60}},\ \bibinfo {pages} {679}
  (\bibinfo {year} {2011})}\BibitemShut {NoStop}%
\bibitem [{\citenamefont {Ortiz}\ \emph {et~al.}(2012)\citenamefont {Ortiz},
  \citenamefont {Cobanera},\ and\ \citenamefont {Nussinov}}]{Ortiz2012}%
  \BibitemOpen
  \bibfield  {author} {\bibinfo {author} {\bibfnamefont {G.}~\bibnamefont
  {Ortiz}}, \bibinfo {author} {\bibfnamefont {E.}~\bibnamefont {Cobanera}}, \
  and\ \bibinfo {author} {\bibfnamefont {Z.}~\bibnamefont {Nussinov}},\ }\href
  {\doibase http://dx.doi.org/10.1016/j.nuclphysb.2011.09.012} {\bibfield
  {journal} {\bibinfo  {journal} {Nuclear Physics B}\ }\textbf {\bibinfo
  {volume} {854}},\ \bibinfo {pages} {780 } (\bibinfo {year}
  {2012})}\BibitemShut {NoStop}%
\bibitem [{\citenamefont {Cook}(1971)}]{Cook1971}%
  \BibitemOpen
  \bibfield  {author} {\bibinfo {author} {\bibfnamefont {S.~A.}\ \bibnamefont
  {Cook}},\ }in\ \href {\doibase 10.1145/800157.805047} {\emph {\bibinfo
  {booktitle} {Proceedings of the third annual ACM symposium on Theory of
  computing}}},\ \bibinfo {series and number} {STOC '71}\ (\bibinfo
  {publisher} {ACM},\ \bibinfo {address} {New York, NY, USA},\ \bibinfo {year}
  {1971})\ pp.\ \bibinfo {pages} {151--158}\BibitemShut {NoStop}%
\bibitem [{\citenamefont {Levin}(1973)}]{Levin1973}%
  \BibitemOpen
  \bibfield  {author} {\bibinfo {author} {\bibfnamefont {L.~A.}\ \bibnamefont
  {Levin}},\ }\href@noop {} {\bibfield  {journal} {\bibinfo  {journal}
  {Problemy Peredachi Informatsii}\ }\textbf {\bibinfo {volume} {9}},\ \bibinfo
  {pages} {115} (\bibinfo {year} {1973})}\BibitemShut {NoStop}%
\bibitem [{\citenamefont {M{\'e}zard}\ and\ \citenamefont
  {Parisi}(1987)}]{Mezard1987}%
  \BibitemOpen
  \bibfield  {author} {\bibinfo {author} {\bibfnamefont {M.}~\bibnamefont
  {M{\'e}zard}}\ and\ \bibinfo {author} {\bibfnamefont {G.}~\bibnamefont
  {Parisi}},\ }\href {http://stacks.iop.org/0295-5075/3/i=10/a=002} {\bibfield
  {journal} {\bibinfo  {journal} {EPL (Europhysics Letters)}\ }\textbf
  {\bibinfo {volume} {3}},\ \bibinfo {pages} {1067} (\bibinfo {year}
  {1987})}\BibitemShut {NoStop}%
\bibitem [{\citenamefont {M{\'e}zard}\ \emph {et~al.}(2002)\citenamefont
  {M{\'e}zard}, \citenamefont {Parisi},\ and\ \citenamefont
  {Zecchina}}]{Mezard2002}%
  \BibitemOpen
  \bibfield  {author} {\bibinfo {author} {\bibfnamefont {M.}~\bibnamefont
  {M{\'e}zard}}, \bibinfo {author} {\bibfnamefont {G.}~\bibnamefont {Parisi}},
  \ and\ \bibinfo {author} {\bibfnamefont {R.}~\bibnamefont {Zecchina}},\
  }\href {\doibase 10.1126/science.1073287} {\bibfield  {journal} {\bibinfo
  {journal} {Science}\ }\textbf {\bibinfo {volume} {297}},\ \bibinfo {pages}
  {812} (\bibinfo {year} {2002})}\BibitemShut {NoStop}%
\bibitem [{\citenamefont {Monasson}\ and\ \citenamefont
  {Zecchina}(1997)}]{Monasson1997}%
  \BibitemOpen
  \bibfield  {author} {\bibinfo {author} {\bibfnamefont {R.}~\bibnamefont
  {Monasson}}\ and\ \bibinfo {author} {\bibfnamefont {R.}~\bibnamefont
  {Zecchina}},\ }\href@noop {} {\bibfield  {journal} {\bibinfo  {journal}
  {Phys. Rev. E}\ }\textbf {\bibinfo {volume} {56}},\ \bibinfo {pages} {1357}
  (\bibinfo {year} {1997})}\BibitemShut {NoStop}%
\bibitem [{\citenamefont {Montanari}\ \emph {et~al.}(2008)\citenamefont
  {Montanari}, \citenamefont {Ricci-Tersenghi},\ and\ \citenamefont
  {Semerjian}}]{Montanari2008}%
  \BibitemOpen
  \bibfield  {author} {\bibinfo {author} {\bibfnamefont {A.}~\bibnamefont
  {Montanari}}, \bibinfo {author} {\bibfnamefont {F.}~\bibnamefont
  {Ricci-Tersenghi}}, \ and\ \bibinfo {author} {\bibfnamefont {G.}~\bibnamefont
  {Semerjian}},\ }\href {\doibase 10.1088/1742-5468/2008/04/P04004} {\bibfield
  {journal} {\bibinfo  {journal} {J. Stat. Mech.}\ }\textbf {\bibinfo {volume}
  {2008}},\ \bibinfo {pages} {04004} (\bibinfo {year} {2008})}\BibitemShut
  {NoStop}%
\bibitem [{Sup()}]{Supplement}%
  \BibitemOpen
  \href@noop {} {}\bibinfo {note}
  {\href{http://terpconnect.umd.edu/~galitski/SATDual/}{\url{http://terpconnect.umd.edu/~galitski/SATDual/}}}\BibitemShut
  {NoStop}%
\bibitem [{\citenamefont {Matveev}\ and\ \citenamefont
  {Shrock}(2008)}]{Matveev2008}%
  \BibitemOpen
  \bibfield  {author} {\bibinfo {author} {\bibfnamefont {V.}~\bibnamefont
  {Matveev}}\ and\ \bibinfo {author} {\bibfnamefont {R.}~\bibnamefont
  {Shrock}},\ }\href {http://stacks.iop.org/1751-8121/41/i=13/a=135002}
  {\bibfield  {journal} {\bibinfo  {journal} {Journal of Physics A:
  Mathematical and Theoretical}\ }\textbf {\bibinfo {volume} {41}},\ \bibinfo
  {pages} {135002} (\bibinfo {year} {2008})}\BibitemShut {NoStop}%
\bibitem [{\citenamefont {Mulders}\ and\ \citenamefont
  {Storjohann}(1999)}]{Mulders1999}%
  \BibitemOpen
  \bibfield  {author} {\bibinfo {author} {\bibfnamefont {T.}~\bibnamefont
  {Mulders}}\ and\ \bibinfo {author} {\bibfnamefont {A.}~\bibnamefont
  {Storjohann}},\ }in\ \href {\doibase 10.1145/309831.309905} {\emph {\bibinfo
  {booktitle} {Proceedings of the 1999 international symposium on Symbolic and
  algebraic computation}}},\ \bibinfo {series and number} {ISSAC '99}\
  (\bibinfo  {publisher} {ACM},\ \bibinfo {address} {New York, NY, USA},\
  \bibinfo {year} {1999})\ pp.\ \bibinfo {pages} {181--188}\BibitemShut
  {NoStop}%
\bibitem [{\citenamefont {J\"{a}ger}\ and\ \citenamefont
  {Wagner}(2009)}]{Jager2009}%
  \BibitemOpen
  \bibfield  {author} {\bibinfo {author} {\bibfnamefont {G.}~\bibnamefont
  {J\"{a}ger}}\ and\ \bibinfo {author} {\bibfnamefont {C.}~\bibnamefont
  {Wagner}},\ }\href {\doibase 10.1016/j.parco.2009.01.003} {\bibfield
  {journal} {\bibinfo  {journal} {Parallel Comput.}\ }\textbf {\bibinfo
  {volume} {35}},\ \bibinfo {pages} {345} (\bibinfo {year} {2009})}\BibitemShut
  {NoStop}%
\bibitem [{\citenamefont {Laumann}\ \emph {et~al.}(2010)\citenamefont
  {Laumann}, \citenamefont {Moessner}, \citenamefont {Scardicchio},\ and\
  \citenamefont {Sondhi}}]{Laumann2010}%
  \BibitemOpen
  \bibfield  {author} {\bibinfo {author} {\bibfnamefont {C.~R.}\ \bibnamefont
  {Laumann}}, \bibinfo {author} {\bibfnamefont {R.}~\bibnamefont {Moessner}},
  \bibinfo {author} {\bibfnamefont {A.}~\bibnamefont {Scardicchio}}, \ and\
  \bibinfo {author} {\bibfnamefont {S.~L.}\ \bibnamefont {Sondhi}},\ }\href
  {http://arxiv.org/abs/0903.1904v1} {\bibfield  {journal} {\bibinfo  {journal}
  {Quant. Inf. and Comp.}\ }\textbf {\bibinfo {volume} {10}},\ \bibinfo {pages}
  {0001} (\bibinfo {year} {2010})}\BibitemShut {NoStop}%
\bibitem [{\citenamefont {Hsu}\ \emph {et~al.}(2013)\citenamefont {Hsu},
  \citenamefont {Laumann}, \citenamefont {Lauechli}, \citenamefont {Moessner},\
  and\ \citenamefont {Sondhi}}]{Hsu2013}%
  \BibitemOpen
  \bibfield  {author} {\bibinfo {author} {\bibfnamefont {B.}~\bibnamefont
  {Hsu}}, \bibinfo {author} {\bibfnamefont {C.~R.}\ \bibnamefont {Laumann}},
  \bibinfo {author} {\bibfnamefont {A.~M.}\ \bibnamefont {Lauechli}}, \bibinfo
  {author} {\bibfnamefont {R.}~\bibnamefont {Moessner}}, \ and\ \bibinfo
  {author} {\bibfnamefont {S.~L.}\ \bibnamefont {Sondhi}},\ }\href@noop {}
  {\bibfield  {journal} {\bibinfo  {journal} {Phys. Rev. A}\ }\textbf {\bibinfo
  {volume} {87}},\ \bibinfo {pages} {062334} (\bibinfo {year}
  {2013})}\BibitemShut {NoStop}%
\bibitem [{\citenamefont {Biere}\ \emph {et~al.}(2009)\citenamefont {Biere},
  \citenamefont {Heule}, \citenamefont {van Maaren},\ and\ \citenamefont
  {Walsh}}]{Biere2009}%
  \BibitemOpen
  \bibinfo {editor} {\bibfnamefont {A.}~\bibnamefont {Biere}}, \bibinfo
  {editor} {\bibfnamefont {M.~J.~H.}\ \bibnamefont {Heule}}, \bibinfo {editor}
  {\bibfnamefont {H.}~\bibnamefont {van Maaren}}, \ and\ \bibinfo {editor}
  {\bibfnamefont {T.}~\bibnamefont {Walsh}},\ eds.,\ \href@noop {} {\emph
  {\bibinfo {title} {Handbook of Satisfiability}}},\ \bibinfo {series}
  {Frontiers in Arificial Intelligence and Applications}, Vol.\ \bibinfo
  {volume} {185}\ (\bibinfo  {publisher} {IOS Press},\ \bibinfo {year} {2009})\
  p.\ \bibinfo {pages} {980}\BibitemShut {NoStop}%
\end{thebibliography}%

%

 \end{document}